\documentclass[11pt, oneside]{article}   	% use "amsart" instead of "article" for AMSLaTeX format
\usepackage{geometry}                		% See geometry.pdf to learn the layout options. There are lots.
\usepackage[parfill]{parskip}    	% Activate to begin paragraphs with an empty line rather than an indent
\usepackage{graphicx}				% Use pdf, png, jpg, or eps§ with pdflatex; use eps in DVI mode
\usepackage{amssymb}
\usepackage{amsmath}
\usepackage[numbers]{natbib}
\usepackage{hyperref}
\usepackage{float}
\usepackage[T1]{fontenc}
\usepackage[utf8]{inputenc}
\usepackage{microtype}
\usepackage{setspace}
\usepackage{lineno}

\begin{document}
\title{An Open-Source Reproduction and Enhancement of CheXNet for Chest X-ray Disease Classification}

\author{
Daniel Strick \and
Carlos Garcia Padilla \and
Anthony Huang \and
Thomas Gardos, PhD
}

\date{
Department of Computing and Data Sciences\\
Boston University\\
665 Commonwealth Ave, Boston, MA 02215
}

\maketitle 
\newpage

\begin{abstract}
Deep learning for radiologic image analysis is a rapidly growing field in biomedical research and is likely to become a standard practice in modern medicine. On the publicly available NIH ChestX-ray14 dataset (also hosted on Kaggle), containing X-ray images that are classified by the presence or absence of 14 different diseases, we reproduced an algorithm known as CheXNet, as well as explored other algorithms that outperform CheXNet’s baseline metrics. Model performance was primarily evaluated using the F1 score and AUC-ROC, both of which are critical metrics for imbalanced, multi-label classification tasks in medical imaging. The best model achieved an average AUC-ROC score of 0.85 and an average F1 score of 0.39 across all 14 disease classifications present in the dataset.
\end{abstract}

\section*{Introduction}

Chest X-ray classification is a crucial task in medical image analysis, where deep learning models are trained to detect various thoracic diseases from radiographic scans. A landmark study in this field, known as CheXNet, introduced a 121-layer DenseNet convolutional neural network that reportedly outperformed radiologists in detecting pneumonia \cite{rajpurkar2017}. Their work used the NIH ChestX-ray14 dataset, a publicly available dataset of over 100,000 frontal-view chest X-rays labeled with up to 14 disease classes \cite{nihxray2017}. The success of CheXNet has inspired further research, as it represents a significant step toward using artificial intelligence to assist in clinical diagnosis, especially in regions where access to licensed radiologists is limited \cite{hwang2023}. In the midst of a reproducibility crisis in academia, independent researchers must reproduce groundbreaking studies like this in order to help guide future research \cite{noorden2015}.
In this project, we set out to replicate the original CheXNet model as closely as possible, evaluate and improve performance metrics such as AUC-ROC and F1 scores across all 14 disease classes, and explore whether newer deep learning techniques, particularly Vision Transformers (ViTs), could offer performance improvements over traditional convolutional neural networks. All code for our models and evaluation pipeline is publicly available in our \href{https://github.com/dstrick17/DacNet}{GitHub Repository - https://github.com/dstrick17/DacNet}. While our primary goal was to replicate the original CheXNet study, we also recognize the importance of its successor, CheXNeXt, which validated a similar model against board-certified radiologists on a curated internal dataset \cite{rajpurkar2018}. Although the test set used in CheXNeXt is not publicly available, its findings emphasize the clinical relevance of these models and reinforce the need for reproducible benchmarking in public datasets like NIH ChestX-ray14.

Our key contributions are as follows:
We performed a faithful replication of the CheXNet model, establishing a reproducible baseline using pretrained DenseNet-121 with standard training procedures. We proposed an improved model, DACNet, which incorporates Focal Loss, the AdamW optimizer, and advanced image augmentations like Color Jitter. It achieved significantly higher F1 scores on rare classes compared to the baseline. We implemented per-class F1 threshold optimization to further boost classification accuracy, especially in multi-label settings. Unlike the original CheXNet study, which only reported an F1 score for pneumonia using a non-public expert-labeled subset, our study computes per-class F1 scores across all 14 diseases using a reproducible patient-wise split. This provides a more granular view of model strengths and limitations in multi-label medical image classification. We explored the use of transformer-based models (ViT) for X-ray classification, benchmarking their performance against CNN-based architectures. Finally, we developed a Streamlit web app hosted on Hugging Face that takes a chest X-ray input, returns disease predictions using DACNet, and overlays Grad-CAM heatmaps to visualize model attention.

\section*{Related Work}

Deep learning methods have been proven to achieve human-like accuracy at detecting images and have the potential to perform far better than human experts. In 2012, a CNN known as AlexNet demonstrated human-like performance in image classification tasks \cite{krizhevsky2012}. Our main source of inspiration comes from the paper “CheXNet: Radiologist-Level Pneumonia Detection on Chest X-Rays with Deep Learning”. The authors in this paper used Deep Learning to accurately diagnose pneumonia as well as 13 other diseases from X-ray images alone. In addition to CheXNet, the authors later introduced CheXNeXt, a related model evaluated directly against radiologists on 14 thoracic diseases. Their results showed radiologist-level performance on most conditions, reinforcing the potential of deep learning for clinical applications. However, the curated validation set used in CheXNeXt is not publicly available, making full replication difficult. Our study complements this work by rigorously benchmarking against public data using patient-wise splits and transparent model evaluation. Despite the substantial advancements Deep Learning has made in medical imaging, there is still room for improvement \cite{shen2017}. This is evidenced by recent publications that have outperformed previous state-of-the-art deep learning models in tasks such as analyzing Chest X-ray abnormalities \cite{kufel2023}.
 Other papers have been published reporting Deep Learning’s efficacy in accurately diagnosing diseases in a variety of medical fields, including ophthalmology \cite{kermany2018}. A study from Imagen Technologies has been published, which demonstrated that a deep learning system can accurately emulate the expertise of orthopedic surgeons and radiologists at detecting fractures in adult musculoskeletal radiographs \cite{jones2020}.

\section*{Approach}

\subsection*{Data}

Our project initially aimed to reproduce the CheXNet study using a 121-layer DenseNet convolutional neural network trained on chest X-rays. This dataset contains over 100{,}000 frontal-view chest X-rays labeled with up to 14 disease classes, making it a robust benchmark for evaluating multi-label classification models in medical imaging.

One major challenge is the extreme class imbalance within the dataset. While there are only 14 possible classifications in our X-ray dataset, any given image can exhibit any combination of these diseases. This creates a theoretical space of $2^{14}$, or 16{,}384 possible label combinations. However, most of these combinations are extremely rare (e.g., an image containing all 14 conditions) and are not included in the dataset.

Exploratory data analysis revealed 836 unique combinations of disease labels. Notably, approximately 50\% of the images are labeled as ``No Finding,'' while 8\% are labeled as ``Infiltration.'' The remaining 800 combinations each account for less than 4\% of the dataset, highlighting the extreme label imbalance and sparsity. This diversity in label combinations presents a substantial challenge for metrics like the F1 score, particularly for rare conditions.
\begin{table}[ht]
\centering
\begin{tabular}{|l|r|r|}
\hline
\textbf{Combination} & \textbf{Number of Cases} & \textbf{Percentage of Total Cases} \\
\hline
No Finding & 60,361 & 53.84\% \\
Infiltration & 9,547 & 8.51\% \\
Atelectasis & 4,215 & 3.76\% \\
Effusion & 3,955 & 3.53\% \\
Nodule & 2,705 & 2.41\% \\
Pneumothorax & 2,194 & 1.96\% \\
Mass & 2,139 & 1.91\% \\
Effusion, Infiltration & 1,603 & 1.43\% \\
Atelectasis, Infiltration & 1,350 & 1.20\% \\
Consolidation & 1,310 & 1.17\% \\
Atelectasis, Effusion & 1,165 & 1.04\% \\
Pleural\_Thickening & 1,126 & 1.00\% \\
Cardiomegaly & 1,093 & 0.97\% \\
Emphysema & 892 & 0.80\% \\
\hline
\end{tabular}
\caption{Most common findings in the NIH ChestX-ray14 dataset}
\label{tab:common_findings}
\end{table}

\subsection*{Set Up}
We also tested several architectural and training modifications. These included replacing the default optimizer with AdamW, adding color jitter to the image transformation pipeline, using a cosine annealing learning rate scheduler, and swapping DenseNet with alternative backbones such as ResNet-50, EfficientNetB3, and a ViT transformer model. After isolating the effects of each modification, only three changes led to consistent performance improvements: using focal loss instead of binary cross-entropy, adopting the AdamW optimizer, and incorporating color jitter. All three enhancements were integrated into our best-performing model.

Additionally, we found that using unique, per-class F1 thresholds significantly boosted performance compared to applying a global threshold across all diseases. This approach allowed the model to more effectively balance precision and recall for each individual disease, especially those with low prevalence.

\subsection*{Architecture}

\begin{enumerate}
    \item \textbf{Replicate\_CheXNet}: A faithful replication of the original CheXNet paper. We trained a pretrained DenseNet-121 with default BCE loss, image resizing to 224x224, random horizontal flip, and an Adam optimizer with a learning rate of 0.001. This model established a performance baseline for our project, achieving reasonable AUCs but low F1 scores, particularly on underrepresented classes like pneumonia. The CheXNet replica model had a test average AUC of 0.79, test loss of 0.17, and a test average F1 score of just 0.08. This model only had a higher AUC score than the original CheXNet paper for 2 out of the 14 diseases.

    \item \textbf{DACNet}:  Our best-performing model. We built on DenseNet-121 but replaced the BCE loss with Focal Loss (gamma=2, alpha=1) to address extreme class imbalance. We used the AdamW optimizer with weight decay, a learning rate of 0.00005, and a ReduceLROnPlateau scheduler. Augmentations included RandomResizedCrop(224), RandomHorizontalFlip, and ColorJitter. This model was trained using a patient-level split and achieved a test AUC of 0.85, a test loss of 0.04, and an average F1 score of 0.39. We believe the focal loss contributed significantly to reducing test loss and improving prediction confidence on minority classes. This model outperformed CheXNet in AUC for 9 out of 14 diseases.

    \item \textbf{ViT\_Transformer}: We also explored a Vision Transformer-based architecture to test whether newer attention-based methods could outperform CNNs. We used the Hugging Face implementation of ViT pre-trained on ImageNet and fine-tuned it on our X-ray dataset. Although ViTs demonstrated promising potential in other tasks, they did not improve performance on this dataset, likely due to the limited number of training images and the need for longer fine-tuning. This model produced an average test AUC of 0.794, a test loss of 0.159, and an average test F1 score of 0.111 across all diseases. This model did not beat CheXNet’s AUC score for any of the 14 diseases.
\end{enumerate}

Lastly, we developed a Streamlit app hosted on Hugging Face that allows users to upload a chest X-ray image and receive disease predictions using our DACNet model. The app also features Grad-CAM visualizations, which generate heatmaps to highlight the regions of the image most influential in the model’s predictions. This functionality can help radiologists and patients better understand the model’s reasoning and identify potential areas of concern within the X-ray. This streamlit app can be found at \url{https://huggingface.co/spaces/cfgpp/DACNet}. 
\begin{figure}[ht]
    \centering
    \includegraphics[width=0.8\textwidth]{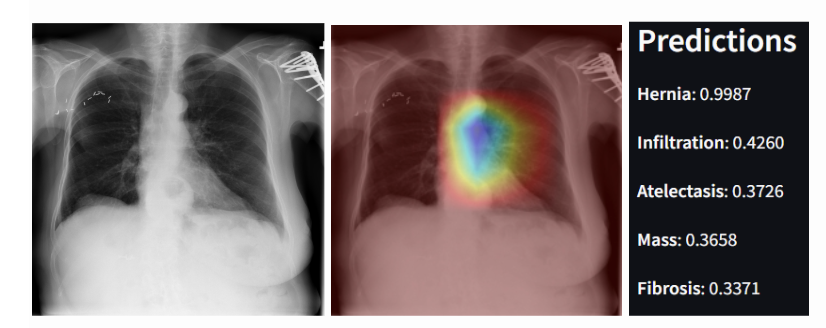} 
    \caption{(Left) An example of an X-Ray from the NIH ChestX-ray14 with Hernia and Infiltration as its ground truth findings. (Center) Grad-CAM visualizations of the same X-Ray image. (Right) DACNet’s predictions of the 5 most likely findings in the image.
}
    \label{fig:yourlabel} 
\end{figure}

This is an example of an X-Ray diagnosed with both a Hernia and Infiltration. The model correctly assigns the highest probability to the correct diseases (hernia and infiltration), but it only predicts that Hernia has a probability above 50\%. This behavior reflects the model's high sensitivity (as indicated by a high AUC) but lower precision, leading to a lower F1 score. The model is effective at ranking the correct disease as the highest, but less effective at distinguishing between the presence and absence of each disease. This observation is consistent with the trade-off between sensitivity and specificity in imbalanced datasets.

\subsection*{Evaluation}

Throughout the study, we explored a wide range of techniques, many of which were ultimately removed from our final model. To address class imbalance, we experimented with oversampling rare disease cases, generating synthetic samples, and tuning the F1 threshold based solely on the rarest disease. While these methods marginally improved scores for targeted diseases, they significantly reduced the average AUC across all conditions, making them ineffective overall.
One surprising outcome came from incorporating demographic data such as age and gender. While these features resulted in a very slight improvement in average test AUC scores, they had no meaningful impact on F1 scores. Given the minimal benefit and increased model complexity, we decided to exclude demographic features from the final model.

\section*{Datasets}

All data used in this study were from the NIH-chestX-ray14 dataset, the same data that was used to build the CheXNet algorithm. This publicly available dataset can be found at \url{https://www.kaggle.com/datasets/nih-chest-xrays/data} \cite{nihxray2017} or at \url{https://nihcc.app.box.com/v/ChestXray-NIHCC} \cite{nihpress2017}. 

\section*{Evaluation Results}

\subsection*{Reproducibility Note}
While we compare our results to those reported by CheXNet and its successor, CheXNeXt, it is important to note a key limitation in reproducibility. The CheXNeXt study evaluated its model on a subset of 420 images from the NIH ChestX-ray14 dataset, but using expert-verified labels provided by four board-certified radiologists. These annotations were not released publicly, making their model-to-radiologist comparison non-reproducible. In contrast, the CheXNet study used the original NIH labels for training and testing on a public patient-wise split. While this allows for reproducible AUC evaluation across all 14 diseases, their F1 score comparison to radiologists also relied on the expert-labeled 420-image subset, which we cannot replicate.
\begin{figure}[H]
    \centering
    \includegraphics[width=0.85\textwidth]{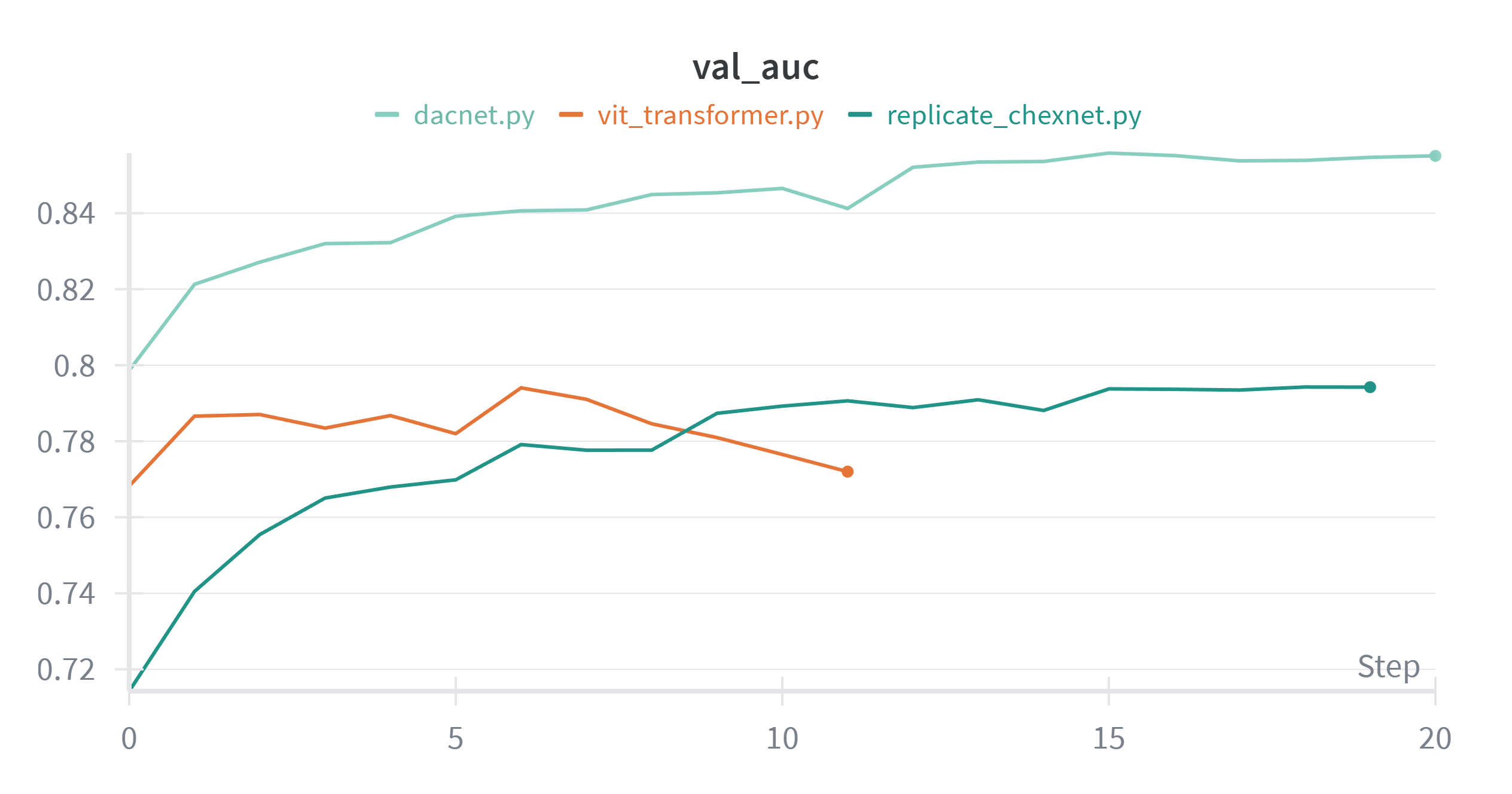} % Replace with your actual filename
    \caption{Graphs from \texttt{wandb.ai} showing comparisons of the three final models: \texttt{DACNet.py}, \texttt{vit\_transformer.py}, and \texttt{replicate\_chexnet.py}. The average validation AUC score across all diseases is shown across the entire training run.}
    \label{fig:wandb_comparison}
\end{figure}

\begin{figure}[H]
    \centering
    \includegraphics[width=0.85\textwidth]{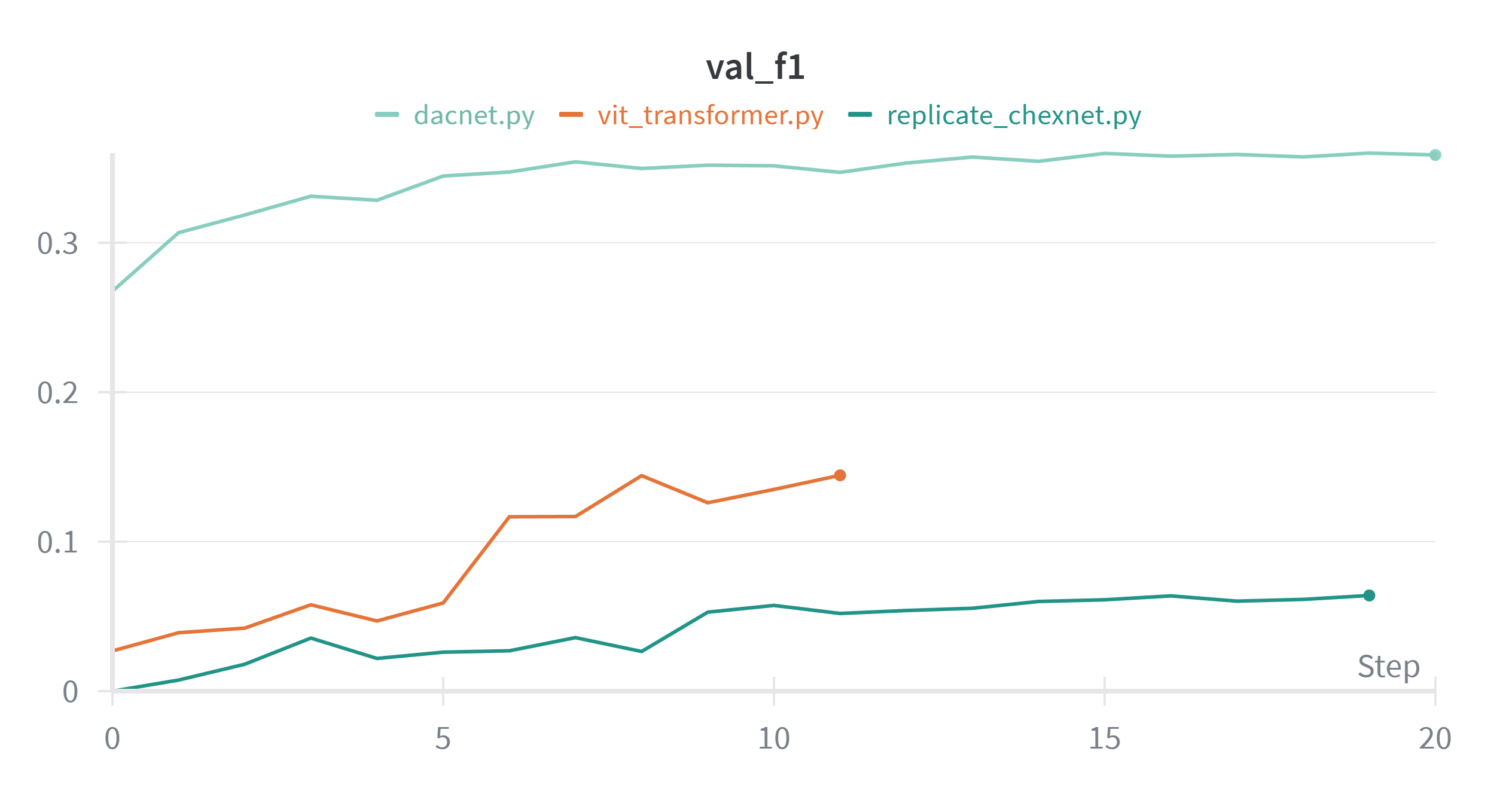} % Replace with actual filename
    \caption{Graphs from \texttt{wandb.ai} showing comparisons of the three final models: \texttt{DACNet.py}, \texttt{vit\_transformer.py}, and \texttt{replicate\_chexnet.py}. The average validation F1 score across all diseases is shown across the entire training run.}
    \label{fig:wandb_f1_comparison}
\end{figure}

\begin{figure}[H]
    \centering
    \includegraphics[width=0.85\textwidth]{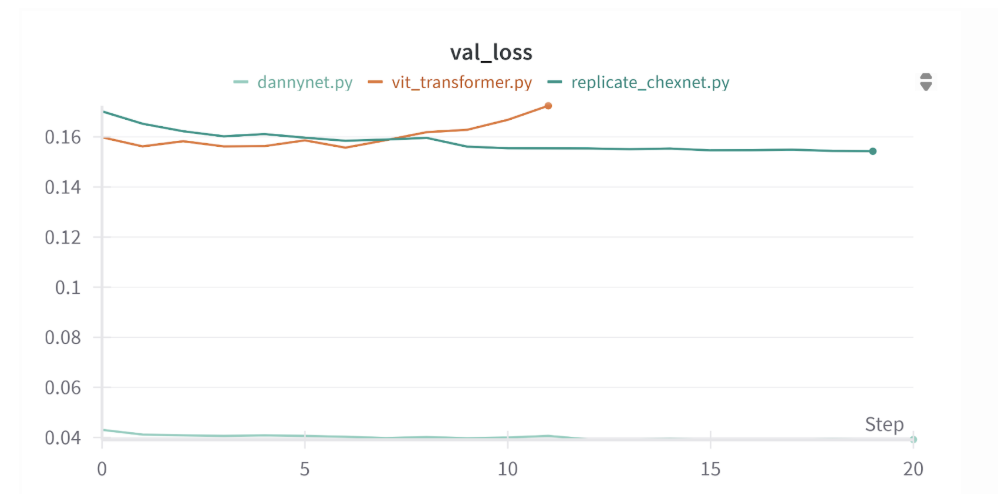} % Replace with actual filename
    \caption{Graphs from \texttt{wandb.ai} showing comparisons of the three final models: \texttt{DACNet.py}, \texttt{vit\_transformer.py}, and \texttt{replicate\_chexnet.py}. The average validation loss across all diseases is shown across the entire training run.}
    \label{fig:wandb_loss_comparison}
\end{figure}

For model evaluation, we shifted our primary focus from F1 score in pneumonia cases which was used in the original CheXNet paper to AUC scores for all 14 disease classes. The original study evaluated F1 for pneumonia on a private dataset that we did not have access to, making exact replication impossible. As a result, we prioritized publicly replicable metrics and broader performance measures.
Through these experiments, we found that simply replicating CheXNet’s structure was not sufficient to achieve high F1 scores, especially for underrepresented diseases. Architectural refinements, the use of more appropriate loss functions, and carefully selected augmentations proved essential. Our final model, DACNet, emerged as a robust and interpretable improvement upon the original pipeline.
DACNet was trained using a stratified patient-level split across train, validation, and test sets to prevent information leakage between patients. It was evaluated across all 14 disease classes using both F1 and AUC. We also integrated Grad-CAM visualizations to improve interpretability and user trust in the model’s predictions.

DACNet achieved a test AUC of approximately 0.85 and an average F1 score of 0.39, which significantly outperformed earlier versions of our model that used basic binary cross-entropy loss and static thresholds. While the inclusion of focal loss had negligible impacts on test AUC and test F1, it led to a marked reduction in test loss, suggesting that the model’s predictions became more confident and better calibrated for rare disease classes.

Of note, only DACnet.py had a custom F1 threshold for each individual disease; replicate\_chexnet.py and transformer.py had a threshold of 0.5 for all diseases, which was likely too high. Since the CheXNet paper did not explicitly mention the threshold, we decided to pick a standard threshold of 0.5.

The Streamlit app demonstrates that the model consistently ranks the correct diagnosis as the highest-probability prediction when that condition is present. However, it often assigns moderate probabilities to additional diseases that are not actually present. This behavior aligns with the model’s high AUC score but relatively low F1 score. A high AUC suggests the model is good at ranking positive examples above negatives, but a low F1 indicates it struggles with making precise binary decisions about whether each disease is truly present or absent. As a result, while the app is effective at highlighting the most likely condition, it is less reliable at cleanly distinguishing between present and absent diseases.

\begin{table}[H]
\centering
\resizebox{\textwidth}{!}{%
\begin{tabular}{|l|c|c|c|c|}
\hline
\textbf{Pathology} & \textbf{Original CheXNet (2017)} & \textbf{DACNet (Ours)} & \textbf{Transformer Model (Ours)} & \textbf{Replicate CheXNet (Ours)} \\
\hline
Atelectasis & 0.809 & \textbf{0.817} & 0.774 & 0.762 \\
Cardiomegaly & 0.925 & \textbf{0.932} & 0.890 & 0.922 \\
Consolidation & \textbf{0.790} & 0.783 & 0.789 & 0.746 \\
Edema & 0.888 & \textbf{0.896} & 0.876 & 0.864 \\
Effusion & 0.864 & \textbf{0.905} & 0.857 & 0.883 \\
Emphysema & 0.937 & \textbf{0.963} & 0.828 & 0.850 \\
Fibrosis & 0.805 & \textbf{0.814} & 0.772 & 0.766 \\
Hernia & 0.916 & \textbf{0.997} & 0.872 & 0.925 \\
Infiltration & \textbf{0.735} & 0.708 & 0.700 & 0.673 \\
Mass & 0.868 & \textbf{0.919} & 0.783 & 0.824 \\
Nodule & 0.780 & \textbf{0.789} & 0.673 & 0.646 \\
Pleural Thickening & \textbf{0.806} & 0.801 & 0.766 & 0.756 \\
Pneumonia & \textbf{0.768} & 0.740 & 0.713 & 0.656 \\
Pneumothorax & \textbf{0.889} & 0.875 & 0.821 & 0.827 \\
\hline
\end{tabular}%
}
\caption{A comparison of the three models built in the study to the original CheXNet model in Test AUC scores for each disease.}
\label{tab:test_auc_scores}
\end{table}

\begin{table}[H]
\centering
\renewcommand{\arraystretch}{1.2}
\resizebox{0.8\textwidth}{!}{%
\begin{tabular}{|l|c|c|c|}
\hline
\textbf{Metric} & \textbf{DACNet} & \textbf{ViT Transformer} & \textbf{Replicate CheXNet} \\
\hline
Loss & \textbf{0.0416} & 0.1589 & 0.1661 \\
AUC  & \textbf{0.8527} & 0.7940 & 0.7928 \\
F1   & \textbf{0.3861} & 0.1114 & 0.0763 \\
\hline
\end{tabular}
}
\caption{A comparison of the new models built in this study, evaluating Test Loss, average test AUC, and average test F1 score.}
\label{tab:average_metrics}
\end{table}

\begin{table}[H]
\centering
\renewcommand{\arraystretch}{1.2}
\resizebox{\textwidth}{!}{%
\begin{tabular}{|l|c|c|c|}
\hline
\textbf{Disease} & \textbf{DACNet} & \textbf{ViT Transformer} & \textbf{Replicate CheXNet} \\
\hline
Atelectasis & \textbf{0.421} & 0.127 & 0.026 \\
Cardiomegaly & \textbf{0.532} & 0.264 & 0.423 \\
Consolidation & \textbf{0.226} & 0 & 0 \\
Edema & \textbf{0.286} & 0.004 & 0 \\
Effusion & \textbf{0.623} & 0.427 & 0.459 \\
Emphysema & \textbf{0.516} & 0.079 & 0 \\
Fibrosis & \textbf{0.127} & 0 & 0 \\
Hernia & \textbf{0.750} & 0 & 0 \\
Infiltration & \textbf{0.395} & 0.193 & 0.061 \\
Mass & \textbf{0.477} & 0.213 & 0.079 \\
Nodule & \textbf{0.352} & 0.041 & 0 \\
Pleural Thickening & \textbf{0.258} & 0 & 0 \\
Pneumonia & \textbf{0.082} & 0 & 0 \\
\hline
\end{tabular}
}
\caption{A comparison of the three models built in the study in Test F1 scores for each disease.}
\label{tab:f1_by_disease}
\end{table}

\section*{Conclusion}
This project aimed to rigorously reproduce and extend CheXNet, a landmark deep learning model in medical imaging. Through our efforts, we demonstrated that meaningful improvements to the original architecture can be achieved by incorporating techniques developed since the paper’s publication. Specifically, we found that the use of Focal Loss, the AdamW optimizer with weight decay, Color Jitter for data augmentation, ReduceLROnPlateau scheduling, and per-disease F1 threshold tuning substantially improved model stability and performance across all 14 thoracic disease classes.
Our model, "DACNet," achieved a strong balance between interpretability and predictive power, reaching an average AUC of 0.85 and an F1 score of 0.39. These results highlight the potential for targeted enhancements to significantly improve performance on imbalanced and clinically relevant datasets like NIH ChestX-ray14.
Beyond the technical contributions, this project underscores the importance of reproducibility in machine learning research. We were motivated not only to validate the claims of a high-impact study, but also to contribute openly to the scientific community. Our full codebase, including model configurations, evaluation tools, and visualization scripts, has been made publicly available in hopes of promoting transparency and enabling further development by others.
Ultimately, this work contributes to both the biomedical and data science communities by showing how modern training strategies can elevate model performance in critical clinical applications. We hope our findings serve as a foundation for future work exploring deep learning in medical diagnostics and for building more accurate, interpretable, and equitable AI tools in healthcare.

\section*{Role of the Funding Source}

This research did not receive any specific grant from funding agencies in the public, commercial, or not-for-profit sectors. The authors declare that study sponsors had no role in the study design; in the collection, analysis, or interpretation of data; in the writing of the manuscript; or in the decision to submit the manuscript for publication. The authors have no conflicts of interest to declare.

\section*{Generative AI Usage Statement}
Some language in this manuscript was improved using generative AI tools to clarify sentence structure and enhance readability. All AI-generated suggestions were critically reviewed and edited by the authors to ensure accuracy and appropriateness.

\bibliographystyle{plainnat}
\sloppy

\end{document}